\documentstyle[11pt,aaspp4]{article}

\tightenlines

\def\deg{\arcdeg}
\def\eg{{\it e.g.,}}
\def\etal{{\it et al.}}
\def\ie{{\it i.e.,}}

\begin{document}

\title{Deep Spectroscopy in the Field of 3C\,212}

\author{Alan Stockton}
\affil{Institute for Astronomy, University of Hawaii, 2680 Woodlawn
 Drive, Honolulu, HI 96822}
\author{Susan E. Ridgway}
\affil{Department of Physics, University of Oxford, Nuclear and Astrophysics
Laboratory, Keble Road, Oxford, OX1\,3RH, UK}

\begin{abstract}
We present the results of longslit and multiaperture spectroscopy of
faint galaxies in the field of the $z=1.049$ quasar 3C\,212. 
We show that an apparently aligned optical feature beyond the NW radio
lobe has a redshift $z=0.928$, and
a similarly aligned feature just beyond the SE radio lobe has
a redshift $z=1.053$, quite close to that of the quasar.
While the NW optical component is extremely well aligned with the radio
jet and has a morphology that is very similar to that of the radio lobe
lying 3\arcsec\ interior to it, the fact that we find three other field
galaxies with closely similar redshifts indicates that it is most likely a member
of an intervening group rather than an unusual example of true alignment
with the radio structure.  In addition, we have found two galaxies (besides the
one near the SE radio lobe) having redshifts close to that of 3C\,212.
We have firm or probable redshifts for 66 out of 82 galaxies we have
observed in this field.  Comparison with the redshift distribution of faint
field galaxies indicates that a large fraction of the remaining 16 galaxies are
likely to be at redshifts $\gtrsim1.4$.
There are at least two low-redshift dwarf starburst
galaxies showing low metal abundance and high ionization in our sample;
such galaxies are likely to be relatively common in very faint samples,
and, under certain conditions, they could be confused with high-redshift
objects.
\end{abstract}

\section{Introduction}

The alignment of the extended optical continuum and emission-line radiation
with the radio axis in powerful radio galaxies at high redshifts (Chambers,
Miley, \& van Breugel \markcite{cha87}1987; McCarthy et al.\ 
\markcite{mcc87}1987) continues to be a vexing
issue in our understanding of the nature and evolution of such objects.
In spite of strong evidence that a significant portion of this alignment
effect in some objects stems from scattering of radiation from a 
hidden quasar nucleus into
our line of sight (di Serego Aligheri, Cimatti, \& Fosbury \markcite{dis94}1994;
Dickinson, Dey, \& Spinrad \markcite{dic96}1996),
there is also evidence that such scattering is by no means the whole story
(Longair, Best, \& R\"ottgering \markcite{lon95}1995;
Stockton, Ridgway, \& Kellogg \markcite{sto96}1996).
We report elsewhere (Ridgway \& Stockton \markcite{rid97}1997; hereafter
Paper 1) the results of an HST and ground-based imaging survey
of a complete sample of $z\sim1$ 3CR quasars and radio galaxies, including
some of the first clear examples of the alignment effect seen in quasars
(in fact, all 5 of the quasars in our sample show some form of alignment
between the optical and radio structure).  However, instead of clarifying
the nature of the alignment effect, these quasars actually complicate the
picture somewhat, in that at least 3 of them show examples of aligned
optical structure that has such close and detailed correspondence with the
radio structure that it almost certainly is due to optical synchrotron
radiation.  We do not see similar types of alignment in our radio-galaxy
sample, presumably because the aligned synchrotron emission is strongly
beamed and the radio jets in the quasars are closer to our line of sight.

One of the quasars showing this type of alignment is 3C\,212.  
Figure
\ref{colorimg}$A$ shows our HST WFPC2 image with 3.6-cm VLA contours superposed
(see also Fig.\ 6 and Fig.\ 15 of Paper 1).
There is
excellent agreement between the 3 nearly stellar optical peaks
($a$, $b$, and $c$)
just NW of the quasar nucleus and the three peaks in the radio jet.
However, additional apparently aligned structure lies just {\it
beyond} both radio lobes, with some morphological evidence for
association:  the optical component $f$ about 3\arcsec\ beyond the NW
radio lobe appears to mimic the latter's structure, and the object $g$
just beyond the SE radio lobe has faint wisps that seem to parallel
the radio contours.  We have previously discussed some possible
interpretations of these two objects and their relation to 3C\,212
(Paper 1); here we revisit this 
discussion to include the results of spectroscopy of these
features and of a sample of other faint objects in the field
of 3C\,212.

\section{Observations and Data Reduction}

We obtained spectroscopic observations of the 3C212 field
with the Low-Resolution Imaging Spectrometer (LRIS) and the Keck I telescope
on 1995 October 19 and 20 (UT) and 1996 February 14 and 15 (UT).
We used the 600 groove mm$^{-1}$, 7500 \AA\ blaze grating for all
observations.  The 1995 observations used a 1\arcsec\
slit and obtained good spectral coverage from 6670 \AA\ to 9300 \AA.  Unresolved
spectral features have FWHM = 4.8 \AA, and the FWHM of the recorded
quasar continuum ranged from 0\farcs86 to 0\farcs98.  The slit was centered
on the quasar at PA $-42\deg$; this angle places all of the aligned features
on the slit (see Fig.\ \ref{colorimg}$A$).  We obtained three 1200 s 
integrations on 1995 Oct 19 (UT) through intermittent thin clouds; the first of 
these was significantly worse than the remaining two (which appear to have 
suffered negligible extinction) and was subsequently discarded.
The following night, we obtained an additional four 1200 s
integrations under photometric conditions.  The objects were moved 
$\sim5$\arcsec\ along the slit between successive exposures.  The 1996
observations were similar, except that most were obtained with 1\farcs4-wide
slitlets punched in an aperture plate, which also targeted faint galaxies
in the field.  We obtained twelve 1200 s integrations
with this configuration at a position angle of $-41\fdg4$.  We also obtained
three 1200 s integrations with a 0\farcs7-wide standard slit at position angle
$-37\fdg5$, aimed specifically at the bluer portion of the NW component,
as well as three 1200 s integrations in each of two additional aperture
plates covering additional galaxies in the field.

The reduction of the CCD frames followed standard procedures, using
flat-field frames from an internal halogen lamp illuminating the
spectrograph cover plate.  Airglow lines were dealt with by first using
the multiple dithered exposures at each configuration to construct a model
of the airglow spectrum, which was then scaled and subtracted from the
individual exposures.  The long-slit spectra were rectified
in the wavelength coordinate from measurements of the spectrum of an
internal Ne source.  Small corrections to the wavelength zero point
were derived from measurements of positions of airglow OH lines.  After
this rectification, any remaining residual from the airglow spectrum was
removed with the IRAF {\it background} task, generally using either a simple
median or a linear fit along the columns.
We next scanned the spectrum with a series of contiguous
1-pixel-wide apertures covering the region of interest, using tasks
in the IRAF {\it apextract} package.  The scanning parameters were
transferred from a trace of the quasar continuum, for which a 6-pixel-wide
(1\farcs3) aperture was used.  This procedure effectively rectifies the spectra
in the spatial direction and results in a set of spectra that are
aligned with each other in both the spatial and wavelength coordinates.
These spectra were calibrated using observations of the spectrophotometric
standard star G191B2B (Massey et al.\ \markcite{mas88}1988; Massey \& Gronwall 
\markcite{mas90}1990).  These same observations allowed us to correct for 
the atmospheric-absorption A and B bands. 

The multiaperture spectra were treated similarly.  Individual regions 
corresponding to each slitlet were extracted and reduced independently.
The wavelength scale and distortion were determined from OH airglow lines.
Except for the slitlet covering 3C\,212 itself and objects $f$ and $g$, 
we did not attempt to rectify these
multiaperture spectra in the spatial coordinate.  Instead, we used the
IRAF {\it apextract} suite of tasks to trace and extract one-dimensional
spectra of the objects.
The flux calibration is only approximate, having been derived from
the longslit observations of the spectrophotometric standards; however,
because we kept our targets close to the centerline of the aperture plate,
our spectral region is similar to that of the longslit spectra, and the
calibration should be fairly accurate.

\section{Results}
\subsection{Spectroscopy of Aligned Components}

Sections of various two-dimensional spectra covering 3C\,212 with slits
placed along the radio axis are shown in Fig.\ \ref{colorimg}$B$--$E$.
There is an emission feature near 7190 \AA\ in the NW aligned 
component ($f$ in Fig.\ \ref{colorimg}$A$); this is by far the strongest feature
in this object over the observed wavelength range.  
The feature is clearly a close doublet, and deconvolution into two Gaussian
components not only gives the correct wavelength separation
for [\ion{O}{2}] $\lambda\lambda$3726,3729 at a redshift of 0.9284,
but it also gives an intensity ratio of 1.5, which is the expected ratio
at low densities.  This identification 
is confirmed by the presence of a much weaker line at 8373 \AA, the expected 
position of H$\gamma$ (Fig.\ \ref{colorimg}$E$).

\begin{figure}
\plotone{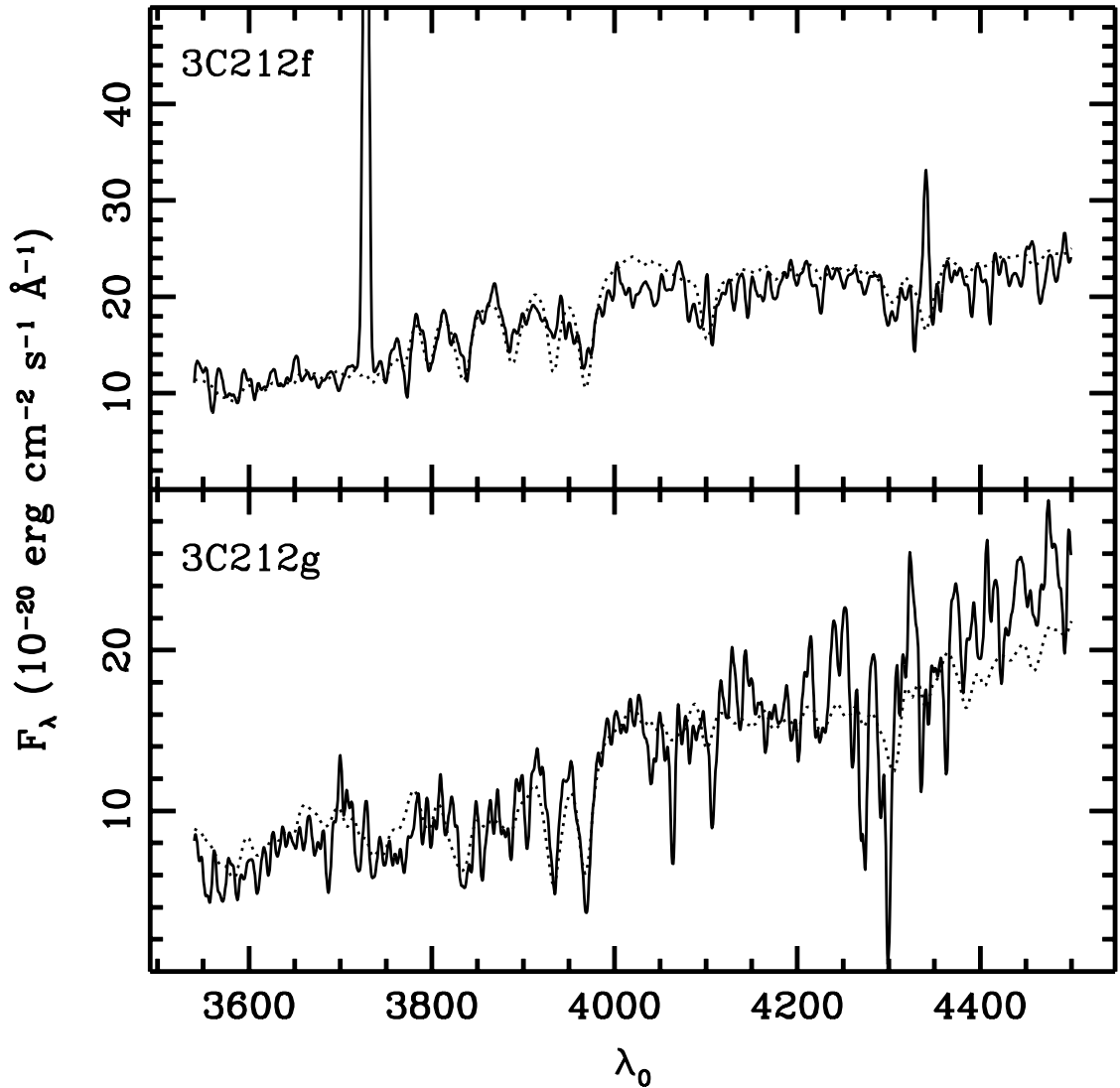}
\caption{Plots of spectra of $f$ and $g$, from the 6-hour combined spectrum
obtained with 1\arcsec\ and 1\farcs4 slits.  Both spectra have been restored
to the rest frame, using $z=0.9283$ for $f$ and $z=1.053$ for $g$.
The dotted traces superposed on the data are of single-epoch solar-metallicity
Bruzual-Charlot 
(1995) isochronal synthesis models:  that for $f$ has an age of
1.0 Gyr, and that for $g$ has an age of 4 Gyr.
}\label{onedspec}
\end{figure}

Figure \ref{onedspec} shows traces of the spectra of the NW and SE aligned 
components beyond the radio lobes ($f$ and $g$).  In addition
to the emission features mentioned above, $f$
shows apparent stellar absorption features.  Both 
the strength of the features
and the continuum shape over the observed wavelength range are well matched 
by a single stellar population with an age of 1 Gyr and solar metallicity
(Bruzual \& Charlot \markcite{bru97}1997).  However, our HST and 
groundbased IR images (Paper 1) 
show that $f$ encompasses a considerable range in color, so a single unreddened
stellar population is clearly an oversimplification.

The SE component $g$ also shows stellar absorption features, but no detectible
emission lines.  The redshift is well determined from the \ion{Ca}{2} 
H and K lines to be 1.053, close to the quasar redshift of 1.0489.  
The continuum color of our spectrum is somewhat
redder even than a 10-Gyr-old Bruzual \& Charlot \markcite{bru97}(1997) 
model (such an age 
would be unrealistic at this redshift), but on an object this faint,
we may be susceptible to small zero-point uncertainties.  Taking into account
both the absorption features and our (restframe) 0.33--1.0 $\mu$m spectral
index from Paper 1, we obtain the best overall agreement with a 4-Gyr
stellar population model having near-solar metallicity.

\subsection{Spectroscopy of Galaxies in the 3C\,212 Field}

The galaxies observed with the aperture plates were selected from
a magnitude-limited ($AB$[7000 \AA]$\le25.3$) sample from a deep image 
obtained in good seeing with the University of Hawaii 88-inch telescope.
The choice of galaxies was controlled partly by our need to obtain deep 
spectroscopy of the aligned components near 3C\,212 itself.  The axis
of our long-exposure aperture plate was at position angle 
$-$41.4, and we chose
galaxies in the available field (7\arcmin\ in the direction perpendicular
to the dispersion) that lay as close as possible to the centerline of
the aperture plate, subject to obtaining efficient packing of slitlets,
each of which was at least 15\arcsec\ long.  We also tried to place the
fainter galaxies on this aperture plate, reserving slightly brighter
objects for two additional aperture plates for which we planned shorter total
exposures.  One of these short-exposure aperture plates was aligned
close to the same angle as the long-exposure plate, and the other was
at roughly 90\deg\ to it (in both cases, after we had chosen the galaxies,
we reoptimized the axes, ending up with position angles of 
$-$42\fdg4 and 53\fdg2).  While our sample of observed galaxies is neither
complete nor unbiased, it should give a reasonable idea of the 
redshifts of major concentrations of galaxies along the line of sight.

\begin{figure}
\plotone{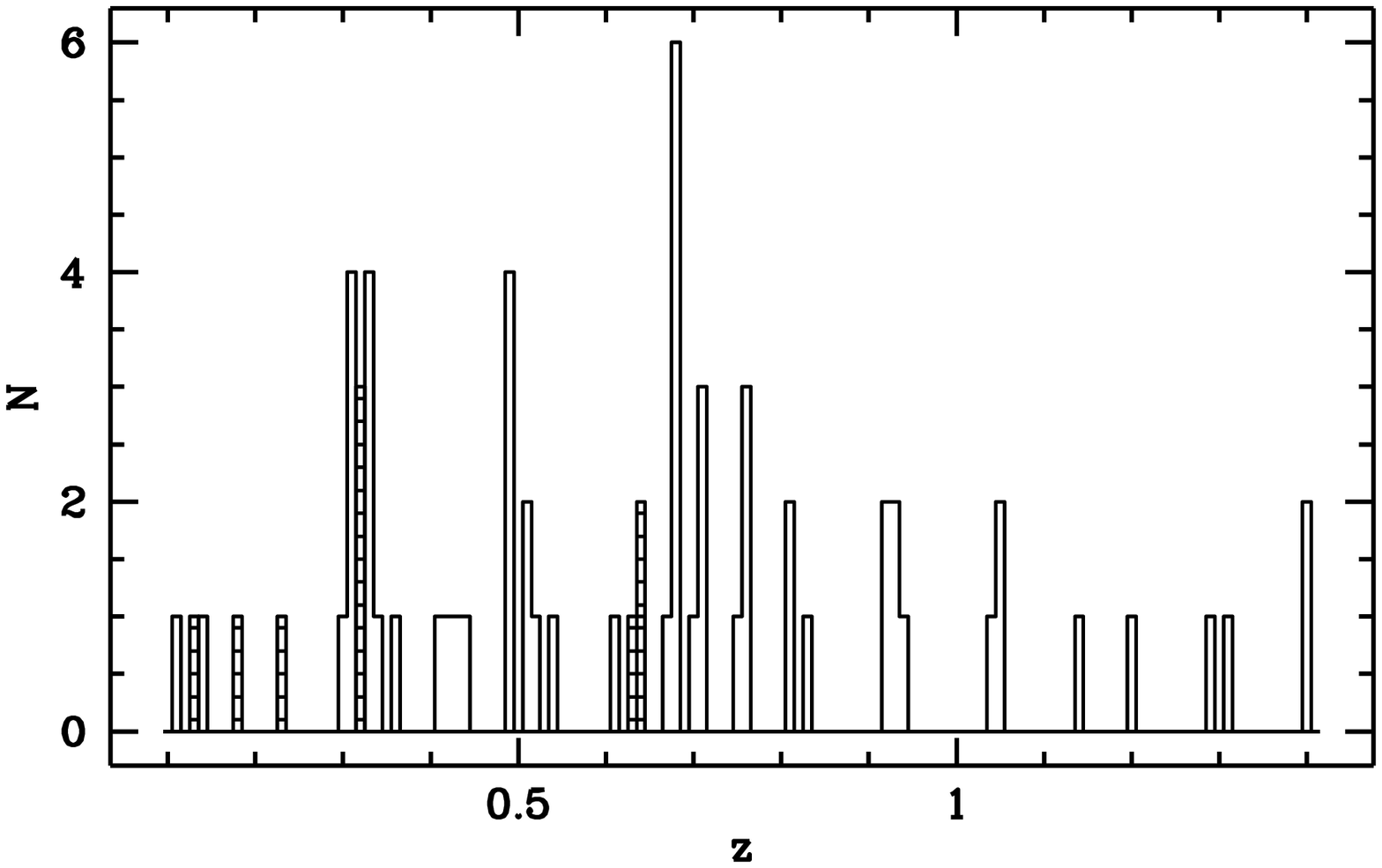}
\caption{Histogram of the redshift distribution of galaxies observed in the
3C\,212 field.
}\label{zhist}
\end{figure}

Table \ref{table1} shows our results for the 82 different objects (besides
the quasar) for which we attempted to obtain spectra.  We have firm or
probable redshifts for 66 of the objects.  Redshifts given in parentheses
are mostly objects for which only one emission line is detected, and in
most cases the line has been identified as H$\alpha$.  In these cases,
we require that the line be both strong and too narrow to be the [\ion{O}{2}] 
$\lambda$3726,3729 doublet, and that the continuum show no evidence for
a discontinuity towards lower levels shortward of the line
(so that it is unlikely to be Ly$\alpha$ at high redshift;
see related discussion in \S\ref{hiigal}).  
Figure \ref{zhist} shows a histogram of the redshift distribution.

\section{Discussion}

\subsection{The Outer Aligned Components of 3C\,212}

We have previously discussed in Paper 1 the problem of the nature of the 
apparently aligned components $f$ and $g$, which lie beyond the radio lobes 
of 3C\,212, and we will only briefly 
summarize the main points of that discussion here before considering the
new data.  The morphological similarity between the optical component $f$ 
and the NW radio lobe of 3C\,212 is truly remarkable (in fact, the initial
reaction of some colleagues to whom we showed these data was that we had 
obviously gotten the scales wrong!).  While $g$ shows less obvious 
structure, it does show faint wisps that appear to extend parallel to
the radio contours, and it lies only about 1\arcsec\ beyond the SW radio
hotspot.

The crux of the matter is that, while there appears to be nearly compelling
morphological evidence for association rather than chance
projection, it is very difficult to understand how any plausible
alignment mechanism can produce strong {\it continuum} sources beyond the
radio lobes.  Furthermore, the radial velocity difference of $\gtrsim18000$
km s$^{-1}$ between the
quasar and $f$ (particularly coupled with the low internal velocity
dispersion of the ionized gas in $f$) adds another constraint on possible
models for association.  The options for $f$ mentioned in Paper 1 were
(1) that it was an unassociated foreground object, (2) that its continuum
was dominated by optical synchrotron emission associated with undetected
radio synchrotron emission (implying a very flat radio---optical spectral
index), (3) that its continuum was dominated by inverse Compton scattering
of microwave background photons associated with undetected or relic
radio radiation (requiring a large total energy in low-energy relativistic
electrons), and (4) that the continuum was due to jet-induced star formation
marking a previous position of the radio jet or a previous outburst.

For $g$, which is only slightly beyond the SE radio lobe, we considered in
Paper 1 the possibility that the faint wisps extending from it could be
thermal emission from a bow shock in the ambient medium beyond the radio
lobe itself.  However, even if this explanation were correct, it cannot
explain the bulk of the emission from $g$, for which both the morphology
and the extremely red color are inconsistent with those expected from 
shock-heated gas.

The new observations we have that are relevant to these questions are
(1) the deeper spectroscopy of $f$ and $g$, which now show stellar absorption
lines in both objects, and (2) the multiaperture spectroscopy of a large
number of field galaxies.  Our detection of stellar absorption lines in
$f$, at the same redshift as the emission lines, effectively eliminates 
any option in which the continuum is
dominated by optical synchrotron or inverse Compton radiation.  It also
eliminates the possibility that we might have been seeing a combination
of emission lines from an intervening object superposed by chance on some form
of continuum at or near the quasar redshift.  The absorption lines in
$g$ finally give us a redshift for this object; the radial velocity
difference of $\sim600$ km s$^{-1}$ in the quasar frame strengthens the case
for association.

In the redshift distribution of galaxies in the 3C\,212 field (Fig.\
\ref{zhist}), there is
clear concentration near $z\sim0.925$, with three objects besides $f$
(B32, C23, and C25) falling within about 600 km s$^{-1}$ of this value 
and one additional object (C05) at $z=0.9435$.  This clustering in redshift 
supports the view
that $f$ is simply a projected intervening galaxy and that the morphological
suggestions of a connection between the radio and optical structure are
fortuitious.  On the other hand, we have found only two additional galaxies
besides $g$ (C17 and C26) having redshifts close to that of the quasar; 
while these galaxies provide evidence for a group around the quasar, there 
is no evidence for a major cluster.

\subsection{The Redshift Distribution of Galaxies in the 3C\,212 Field}

The redshift distribution we observe may be biased by two principal factors:
firstly, we selected galaxies with brighter magnitudes for our two 
shorter-exposure aperture plates than for our long-exposure aperture plate,
so we do not have a consistent magnitude limit for the whole sample; secondly,
our restricted observed wavelength region and differences in the system 
efficiency within this region mean that we will be more sensitive to some
redshift ranges than to others.
Here we estimate the effects of these biases and discuss the likely nature of
the galaxies for which we were not able to obtain redshifts.
We first take the ratio of the number of galaxies in our observed sample in each
one-magnitude interval to that of our total magnitude-limited sample in the
same magnitude interval.  We can then look at the observed redshift 
distribution for galaxies in each magnitude interval and make a correction to
the total redshift distribution, based on normalizing the distribution of 
observed galaxies in the magnitude interval to the number in the same 
interval in the total sample.  
We then renormalize the resulting distribution
back to the total number of our observed sample, rounding to the nearest
unit in each bin.  Figure \ref{zcomp}$a$ shows the 
observed and corrected distributions for $\Delta z=0.1$ bins; the two
distributions are statistically virtually identical.  We now wish to
compare our redshift distribution with that found for field galaxies in a 
large, deep, essentially complete, spectroscopic survey.  The published 
survey best suited for this comparison is that of the 22$^{\rm h}$ field 
of Cowie \etal\ \markcite{cow96}(1996; hereinafter ``CSHC22 sample''), 
which is nearly complete to $I=23$.  Our survey limit of $AB_{7000}=25.3$
corresponds to $I\approx24.4$, assuming spectra that are roughly flat in
$f_{\lambda}$; but the redshift distribution of our fainter sources
is not significantly different from that at $AB_{7000}=24$, so we can use our
whole sample in the comparison.  Figure \ref{zcomp}$b$ shows the redshift
distribution for the CSHC22 sample.  This sample
comprises 167 galaxies, all but 18 of which have firm redshifts.

The most obvious difference (aside from expected clustering peaks) in the 
distributions is the presence of galaxies
with redshifts nearly up to 2.5 in the CSHC22 sample, whereas no
redshifts higher than $\sim1.4$ are seen in our 3C\,212 field redshift 
distribution.
This result is also indicated by a Mann-Whitney $U$ test of the two
samples:  when the 3C\,212 field observed sample is compared with the
whole CSHC22 sample, the probability that the two are drawn
from the same population is about 8\%, whereas, if we delete galaxies
with $z>1.4$ from the latter, the probability rises to $>40$\%.  This
difference has a natural explanation in terms of the restricted wavelength
coverage of our observing configuration.
While the wavelength range of our aperture-plate spectra varied somewhat,
depending on the exact position of the aperture, both the rapid
decline of the CCD response and the increasing strength of the OH airglow
emission work to decrease our detection efficiency at longer wavelengths.
In only one case did we identify a line beyond 9200 \AA\ (H$\alpha$ in B31).
A galaxy for which [\ion{O}{2}] $\lambda3727$ falls at 9200 \AA\ has $z=1.47$.
The next reasonably strong expected emission feature is \ion{C}{3}] 
$\lambda1909$, which falls below our short-wavelength cutoff for 
$z\lesssim2.5$.

This insensitivity to galaxies with redshifts between $\sim1.47$ and
$\sim2.5$ means that it is quite likely that a significant fraction of the 
objects for which we were unable to determine redshifts lie within
this range.  Galaxies with $z>1.4$ constitute 10\% of the
CSHC22 sample for which redshifts have been determined;
in our sample, we lack redshifts for 20\% of our total, so roughly half
(and perhaps more) of these are likely to be high-redshift objects.

\subsection{High-Ionization H\,II Dwarfs}\label{hiigal}

One of our faint field galaxies, B08, shows a spectrum dominated by a 
single strong emission line at 8567 \AA\ on a very weak continuum.
The observed equivalent width (EW) of the emission
line is $\sim640$ \AA, and no other features having
fluxes greater than $\sim2$\% of this line are detected over an observed
wavelength range from 6855 to 9455 \AA.  When a single strong
emission line is seen in a low-resolution, wide-bandwidth, optical
spectrum of an extragalactic object, it is almost certainly likely to be 
one of [\ion{O}{2}] $\lambda$3727, Ly$\alpha$, or possibly H$\alpha$.
Other potential identifications either will normally show other nearby
lines of reasonable strength or are unlikely to have large equivalent widths.
In our spectrum of B08, the line is sufficiently narrow and our resolution
sufficiently high that we can eliminate [\ion{O}{2}] $\lambda$3727 as
a possibility:  we would have resolved the doublet quite easily.  

If the observed line were Ly$\alpha$, the redshift would be 6.05.
However, quite aside from the unprecedented redshift, there is a serious 
problem with such an identification.  Although
the observed continuum is very weak, careful measurements in wavelength
intervals free of strong airglow emission shows that there is no significant
drop in the continuum flux density on the shortward side of the line, as 
would be expected from Ly$\alpha$ forest absorption.  One cannot avoid the 
problem by supposing that many of the clouds might be essentially wholly 
ionized at $z\sim6$, as the flux density ($F_{\lambda}$) is still at close to
the same level at a wavelength where Ly$\alpha$ has $z=4.6$.

This almost insurmountable difficulty with an identification with Ly$\alpha$
leads us to consider more closely the possibility that the line is H$\alpha$.  
While our use of a moderately high dispersion helped us eliminate [\ion{O}{2}]
as a viable identification, here we are hurt by the fact that our restricted
wavelength coverage does not include the expected H$\beta$ region.
The line's large equivalent width
(490 \AA\ in the rest frame, if H$\alpha$) and an upper limit to
[\ion{N}{2}] of $<1$\% of the strength of H$\alpha$ would seem, at first
sight, to argue against this identification.  The [\ion{S}{2}]
$\lambda$6716,6731 lines would lie in a region of strong airglow OH
emission, so the upper limit on their strengths is $\sim2$\% that of
the putative H$\alpha$.  Nevertheless,
models for starbursts (Leitherer \& Heckman \markcite{lei95}1995), which
have been found to be generally consistent with observations
(Stasi\' nska \& Leitherer \markcite{sta96}1996), indicate that H$\alpha$
EWs can be as high as 3000 for ages up to $3\times10^6$ years and can 
remain above $\sim300$ up to ages of $10^8$ years 
(these models assume continuous star formation
with a Saltpeter luminosity function over a mass range from 0.1 to 100
$M_{\sun}$).  Furthermore, there are examples of extreme H\,II galaxies
in which metal abundances are low and very hot stars ($>$60000 K) are
present, suppressing low-ionization species like [\ion{N}{2}] and [\ion{S}{2}]
(\eg\ Tol 1214$-$277; {\it cf.} Fig.\ 4o in Terlevich \etal\
\markcite{ter91}1991).  Such galaxies tend to show weak emission in
\ion{He}{1} $\lambda$5876.  This line would fall within our bandpass, and
we have a marginal detection, which appears to be confirmed by a careful 
examination of the
two-dimensional spectrum, at about 2\% of the strength of the strong
line at 8567 \AA.  We are therefore fairly confident that the latter
actually is H$\alpha$, and this confidence is increased by the presence of
another object, D12, in our sample, which also has
a very weak continuum and strong lines; it would have caused similar
problems of identification had it not had a sufficiently high redshift
that the [\ion{O}{3}] lines fall within our bandpass.  Because such dwarfs 
dominated by very young, metal-poor
populations and showing high ionizations appear not to be too uncommon, it
is clear that seeing a single, strong line with a large equivalent width
is not sufficient for an identification with Ly$\alpha$.  Fortunately, most
spectroscopic surveys of faint galaxies have wider wavelength coverage
than we had and will normally observe the expected [\ion{O}{3}] region
for any lines that might be H$\alpha$ candidates.  The strongest evidence
in favor of Ly$\alpha$ for lines in this region of the spectrum 
(\ie\ redward of $\sim6600$ \AA) would be
a clear indication of a continuum discontinuity across the emission line.

\section{Summary}

We have obtained deep spectroscopy of the two objects $f$ and $g$, which lie 
respectively just beyond the NW and SE radio lobes of 3C\,212.  Object $f$
shows both emission lines and absorption lines from a moderately-young
stellar population at a redshift of 0.9284, blueshifted by some 18,000 
km s$^{-1}$ relative to the quasar.  Object $g$ has the spectrum of an
old ($\gtrsim4$ Gyr) stellar population at a redshift close to that of the
quasar and is likely associated with it.  Our multiaperture spectroscopy of
82 objects in the 3C\,212 field has turned up 2 additional galaxies with
redshifts close to that of 3C\,212, both at fairly large angular separations
from the quasar, as well as 3 galaxies having redshifts close to that of
$f$.  At both of these redshifts, the galaxies all lie within a projected
distance of about 1 Mpc of
3C\,212 and of $f$ ($H_0=75$ km s$^{-1}$ Mpc$^{-1}$, $q_0=1/2$).
Thus, it appears that both 3C\,212 and $f$ are members of rather
loose groups; more importantly, in spite its morphological similarity to
the NW radio lobe of 3C\,212 and its precise alignment with the radio
jet, $f$ is probably an unrelated foreground object.

The redshift distribution of faint galaxies in the 3C\,212 field agrees
well with that of the Cowie \etal\ \markcite{cow96}(1996) 22$^{\rm h}$ 
field except for a
lack of galaxies with $z\gtrsim1.4$ in the former.  This difference is
readily understood in terms of our lack of coverage of wavelength regions
in which any strong emission lines will fall for $1.47\lesssim z\lesssim2.5$,
and it suggests that a large fraction of our objects without redshifts lie
within this range.

The presence of two young dwarf starburst galaxies with high ionizations 
among the fainter members of our sample indicates that such objects
are likely to be fairly common in field-galaxy samples at $R\gtrsim24$.

\acknowledgments
We thank Len Cowie and Esther Hu for discussions and for providing a
machine-readable version of Table 1 from Cowie \etal\ (1996).  We
have enjoyed discussions of the alignment effect in 3C\,212 with
many people, including Neil Trentham, Mark Lacy, and Ken Chambers.
We thank Zachery Ortogero for helping with some of the figures.  This
research was partially supported by NASA through Grant No.\ GO-05401.01-93A
from the Space Telescope Science Institute, which is operated by AURA, Inc.,
under NASA Contract No.\ NAS 5-26555, and by NSF under grant AST95-29078.
SER was supported by a PPARC Postdoctoral Fellowship.

\newpage
\def\oii{[O\,II]}
\def\oiii{[O\,III]}
\def\hbeta{H$\beta$}
\def\halpha{H$\alpha$}
\def\nii{[N\,II]}
\def\neiii{[Ne\,III]}
\def\caii{Ca\,II}
\def\sii{[S\,II]}

\begin{deluxetable}{c c c c c l }
\scriptsize
\tablenum{1}
\tablewidth{350.0pt}
\tablecaption{Spectroscopy of 3C\,212 Field Objects}
\tablehead{
\colhead{Object} & \colhead{$\Delta\alpha$\tablenotemark{a}} & 
\colhead{$\Delta\delta$\tablenotemark{a}} & 
\colhead{$AB_{7000}$\tablenotemark{b}} & \colhead{$z$} &
\colhead{Notes\tablenotemark{c}}}

\startdata

B02 & $-$109\farcs5 & \phs105\farcs9 & 24.8 & 1.311\phn & \oii, \caii, CN \nl
B03 & \phn$-$88\farcs7 & \phs108\farcs2 & 24.5 & 0.6671 & \hbeta, \oiii \nl
B04 & \phn$-$94\farcs8 & \phn\phs83\farcs3 & 25.2 & \nodata & FBC \nl
B05 & \phn$-$75\farcs0 & \phn\phs79\farcs8 & 24.6 & 1.1386 & \oii, \neiii \nl
B06 & \phn$-$63\farcs8 & \phn\phs68\farcs9 & 24.6 & \nodata & NDC \nl
B07 & \phn$-$64\farcs7 & \phn\phs54\farcs0 & 24.2 & \nodata & FBC \nl
B08 & \phn$-$40\farcs9 & \phn\phs57\farcs8 & 24.9 & 0.3053 & \halpha, He\,I $\lambda$5876 \nl
B09 & \phn$-$30\farcs5 & \phn\phs49\farcs6 & 24.9 & 1.402\phn & \oii, \neiii \nl
B10 & \phn$-$30\farcs1 & \phn\phs35\farcs2 & 24.8 & 0.6052 & \hbeta, \oiii \nl
B11 & \phn$-$13\farcs5 & \phn\phs29\farcs3 & 23.7 & \nodata & em. line at 9170.1 \AA \nl
B12 & \phn\phn$-$7\farcs6 & \phn\phs18\farcs9 & 23.6 & 0.3380 & \halpha, \nii \nl
B13 & \phn\phn$-$5\farcs2 & \phn\phn\phs6\farcs3 & 22.6 & 0.9284 & $f$; \oii, H$\gamma$ (see text) \nl
B14 & \phn\phn\phs0\farcs0 & \phn\phn\phs0\farcs0 & 19.0 & 1.0489 & 3C\,212 \nl
B15 & \phn\phn\phs3\farcs9 & \phn\phn$-$4\farcs2 & 25.3 & 1.053\phn & $g$; \caii\ (see text) \nl
B16 & \phn\phs33\farcs3 & \phn\phn\phs2\farcs3 & 24.6 & 1.293\phn & \caii \nl
B17 & \phn\phn\phs1\farcs3 & \phn$-$35\farcs3 & 24.1 & \nodata & FBC \nl
B18 & \phn\phs57\farcs4 & \phn\phn$-$8\farcs4 & 24.8 & \nodata & NDC \nl
B19 & \phn\phs56\farcs4 & \phn$-$37\farcs0 & 23.8 & \nodata & Continuum contaminated by star \nl
B20 & \phn\phs75\farcs5 & \phn$-$37\farcs8 & 24.9 & 0.4881 & \hbeta, \oiii \nl
B21 & \phn\phs89\farcs0 & \phn$-$41\farcs8 & 24.9 & (0.1347) & \halpha \nl
B22 & \phn\phs78\farcs0 & \phn$-$67\farcs5 & 25.3 & 0.4880 & \oiii \nl
B23 & \phn\phs77\farcs7 & \phn$-$82\farcs7 & 24.5 & 0.7144 & \hbeta, \oiii \nl
B24 & \phn\phs93\farcs7 & $-$89\farcs2 & 23.8 & 1.195\phn & \oii, Balmer abs. \nl
B25 & \phn\phs86\farcs5 & $-$115\farcs0 & 24.8 & 0.4307 & \hbeta, \oiii \nl
B26 & \phs110\farcs4 & $-$102\farcs2 & 23.4 & 0.6842 & \hbeta, \oiii, H$\gamma$ \nl
B27 & \phs112\farcs0 & $-$104\farcs3 & 24.4 & 0.6831 & \hbeta, H$\gamma$ \nl
B28 & \phs114\farcs6 & $-$107\farcs5 & 23.3 & 0.6841 & \hbeta, \oiii, H$\gamma$ \nl
B29 & \phs132\farcs1 & $-$109\farcs1 & 24.2 & 0.7569 & \hbeta, \oiii \nl
B30 & \phs113\farcs6 & $-$144\farcs2 & 23.7 & 0.2995 & \halpha, \sii \nl
B31 & \phs148\farcs5 & $-$137\farcs6 & 24.3 & 0.442\phn & \halpha, \hbeta \nl
B32 & \phs120\farcs7 & $-$180\farcs4 & 23.1 & 0.9290 & \oii \nl
B33 & \phs138\farcs8 & $-$187\farcs8 & 24.9 & \nodata & FBC \nl
B34 & \phs134\farcs7 & $-$203\farcs1 & 24.8 & 0.6434 & \hbeta, \oiii \nl
B35 & \phs185\farcs6 & $-$179\farcs0 & 23.8 & (0.2267) & \halpha \nl
C02 & \phs106\farcs9 & \phn\phs98\farcs6 & 23.6 & \nodata & FBC \nl
C03 & \phs111\farcs1 & \phn\phs63\farcs2 & 22.9 & \nodata & FBC \nl
C04 & \phs103\farcs7 & \phn\phs57\farcs4 & 23.9 & (0.3150) & \halpha, poss.\ \nii \nl
C05 & \phn\phs52\farcs9 & \phn\phs87\farcs7 & 22.9 & 0.9435 & \oii, \neiii \nl
C06 & \phn\phs74\farcs0 & \phn\phs35\farcs8 & 23.8 & 0.7614 & \hbeta, \oiii \nl
C07 & \phn\phs44\farcs1 & \phn\phs36\farcs4 & 22.3 & 0.3136 & \halpha, \nii \nl
C08 & \phn\phs33\farcs9 & \phn\phs27\farcs9 & 22.8 & 0.3323 & \halpha, \nii \nl
C09 & \phn\phs31\farcs6 & \phn\phs13\farcs9 & 22.9 & 0.7032 & \hbeta, \oiii \nl
C10 & \phn\phs21\farcs1 & \phn\phn\phs3\farcs3 & 23.1 & 0.4921 & \hbeta, \oiii \nl
C11 & \phn\phn$-$5\farcs2 & \phn\phn\phs6\farcs3 & 22.6 & 0.9283 & $f$ (same as
B13); \oii \nl
C12 & \phn$-$13\farcs8 & \phn\phn$-$4\farcs5 & 23.0 & \nodata & FBC \nl
C13 & \phn$-$32\farcs9 & \phn$-$14\farcs5 & 23.0 & 0.5167 & \oiii \nl
C14 & \phn$-$53\farcs4 & \phn$-$21\farcs9 & 22.0 & 0.4884 & \hbeta, \oiii \nl
C15 & \phn$-$61\farcs8 & \phn$-$34\farcs3 & 23.3 & 0.4201 & \oiii \nl
C16 & \phn$-$56\farcs3 & \phn$-$58\farcs5 & 23.0 & \nodata & FBC \nl
C17 & \phn$-$80\farcs6 & \phn$-$54\farcs1 & 22.9 & 1.0505 & \caii, G-band \nl
C18 & \phn$-$79\farcs7 & \phn$-$79\farcs9 & 21.2 & 0.3056 & \halpha, \nii, \sii \nl
C19 & $-$114\farcs5 & \phn$-$52\farcs8 & 22.2 & 0.5100 & \hbeta, \oiii \nl
C20 & $-$114\farcs3 & \phn$-$86\farcs7 & 22.6 & 0.8070 & \caii \nl
C21 & $-$128\farcs1 & \phn$-$89\farcs5 & 23.7 & 1.397\phn & \oii \nl
C22 & $-$135\farcs0 & $-$121\farcs0 & 22.7 & 0.8318 & \hbeta, \oiii \nl
C23 & $-$141\farcs4 & $-$125\farcs5 & 22.1 & 0.9221 & \oii \nl
C24 & $-$156\farcs6 & $-$122\farcs4 & 22.5 & 0.7575 & \hbeta, \oiii \nl
C25 & $-$168\farcs3 & $-$123\farcs8 & 21.4 & \phd0.924:\phn & 4000 \AA\ break \nl
C26 & $-$175\farcs1 & $-$154\farcs2 & 22.3 & 1.043\phn & \oii, \neiii, poss.\ H$\gamma$ \nl
C27 & $-$202\farcs9 & $-$144\farcs9 & 22.6 & 0.7108 & \hbeta, \oiii \nl
D01 & $-$123\farcs5 & \phn\phs88\farcs4 & 23.2 & (0.1790) & \halpha \nl
D02 & \phn$-$65\farcs4 & \phs111\farcs6 & 22.8 & \nodata & FBC \nl
D03 & \phn$-$78\farcs2 & \phn\phs77\farcs1 & 22.9 & (0.6256) & \hbeta, poss.\ \oiii, H$\gamma$ \nl
D04 & \phn$-$74\farcs8 & \phn\phs60\farcs3 & 23.5 & \nodata & FBC \nl
D05 & \phn$-$64\farcs7 & \phn\phs41\farcs5 & 20.3 & 0.3294 & \halpha, \nii, Na D \nl
D06 & \phn$-$72\farcs9 & \phn\phs11\farcs9 & 22.7 & 0.000\phn & M4 dwarf star \nl
D07 & \phn$-$33\farcs7 & \phn\phs26\farcs3 & 23.0 & 0.3051 & \halpha, \nii \nl
D08 & \phn\phn$-$4\farcs4 & \phn\phs37\farcs4 & 21.2 & 0.3301 & \halpha, \nii \nl
D09 & \phn$-$28\farcs1 & \phn\phn$-$4\farcs1 & 21.1 & 0.3331 & \halpha, \sii \nl
D10 & \phn$-$15\farcs1 & \phn$-$12\farcs5 & 23.8 & (0.136)\phn & \halpha \nl
D11 & \phn$-$11\farcs2 & \phn$-$16\farcs8 & 22.5 & 0.8091 & H$\epsilon$ and higher Balmer abs. \nl
D12 & \phn\phs11\farcs1 & \phn$-$21\farcs3 & 23.8 & 0.4139 & \oiii, \halpha \nl
D13 & \phn\phs50\farcs2 & \phn$-$22\farcs4 & 22.0 & 0.5365 & \hbeta, \oiii \nl
D14 & \phn\phs71\farcs6 & \phn$-$39\farcs5 & 22.4 & 0.5086 & \hbeta, \oiii \nl
D15 & \phn\phs69\farcs1 & \phn$-$73\farcs0 & 22.6 & 0.1140 & \halpha, \sii \nl
D16 & \phs100\farcs8 & \phn$-$69\farcs5 & 22.7 & 0.7147 & \hbeta, \oiii \nl
D17 & \phn\phs86\farcs4 & \phn$-$98\farcs6 & 23.8 & (0.3179) & \halpha \nl
D18 & \phn\phs90\farcs3 & $-$111\farcs5 & 21.6 & 0.6834 & \hbeta, H$\gamma$ \nl
D19 & \phs114\farcs0 & $-$113\farcs6 & 22.7 & 0.6837 & \hbeta, H$\gamma$ \nl
D20 & \phn\phs99\farcs9 & $-$142\farcs3 & 23.4 & \nodata & FBC \nl
D21 & \phs105\farcs4 & $-$150\farcs5 & 23.1 & \nodata & poss.\ em.\ line at 8302.4 \AA \nl
D22 & \phs143\farcs3 & $-$137\farcs0 & 22.8 & 0.3560 & \halpha, \nii \nl
D23 & \phs169\farcs3 & $-$142\farcs5 & 22.6 & (0.6439) & \oiii \nl
D24 & \phs133\farcs8 & $-$188\farcs9 & 23.4 & 0.6819 & \hbeta, \oiii, H$\gamma$ \nl
D25 & \phs140\farcs1 & $-$195\farcs1 & 23.0 & 0.7545 & \hbeta, \oiii \nl
\enddata
\tablenotetext{a}{Coordinate offsets from 3C\,212.  These should be accurate
to $\sim0\farcs3$.
}
\tablenotetext{b} {$AB$ magnitude in a 4\arcsec\ diameter aperture, from a 
bandpass centered at 7000 \AA\ and having a FWHM of 2000 \AA.  A 1-$\sigma$
detection above sky would correspond to $AB_{7000}=25.8$.
}
\tablenotetext{c} {Includes spectral features on which the redshift is based; 
however, not all of these features have necessarily been used to determine the 
value of the redshift.  Emission features:  \oii\ $\lambda\lambda3726$,3729,
\neiii\ $\lambda3868$, H$\gamma$ $\lambda4340$, He\,I $\lambda5876$, 
\hbeta\ $\lambda4861$, \oiii\ $\lambda\lambda4959$,5007, \halpha\ $\lambda6563$,
\nii\ $\lambda6583$, \sii\ $\lambda\lambda6716$,6731.  Absorption features:
H$\epsilon$ and higher Balmer lines, \caii\ H and K, Na\,I D.  Other
abbreviations:  FBC = faint blue continuum; NDC = no detected continuum.
}
\tablecomments{Redshifts in parentheses depend on only one line (usually
identified as \halpha) or are otherwise uncertain.}\label{table1}
\end{deluxetable}
\def\deg{\arcdeg}
\pagestyle{empty}
\setlength{\headheight}{0cm}
\setlength{\headsep}{0cm}
\def\etal{{\it et al.}}

\begin{figure}[p]
\caption{{\it HST} image and spectra of objects near 3C\,212.  {\it (A)} HST
image, aligned so that position angle $-$41\fdg4 is vertical.  Object
designations are the same as in Ridgway \& Stockton (1997).  The 3.6-cm
VLA map is overlain as green contours.  Tickmarks are at 1\arcsec\ 
intervals.  {\it (B--D)} Keck LRIS spectral maps of the region of the [O\,II]
$\lambda$3727 doublet in object $f$.  The dispersion direction is
horizontal, and the vertical spatial coordinate is aligned with that of
image {\it (A)}.  {\it (B)}  Combination of 4 hours of data with a
1\farcs4 slit at PA $-$41\fdg4 and 2 hours of data with a 1\arcsec\ slit
at PA $-$42\deg; {\it (C)} 2 hours with a 1\arcsec\ slit at PA $-$42\deg\
alone; {\it (D)} 1 hour with a 0\farcs7 slit at PA $-$37\fdg5.  
{\it (E)} From the same combined spectrum as {\it (B)}, but showing the
H$\gamma$ line.
}\label{colorimg}
\end{figure}

\begin{figure}
\plotone{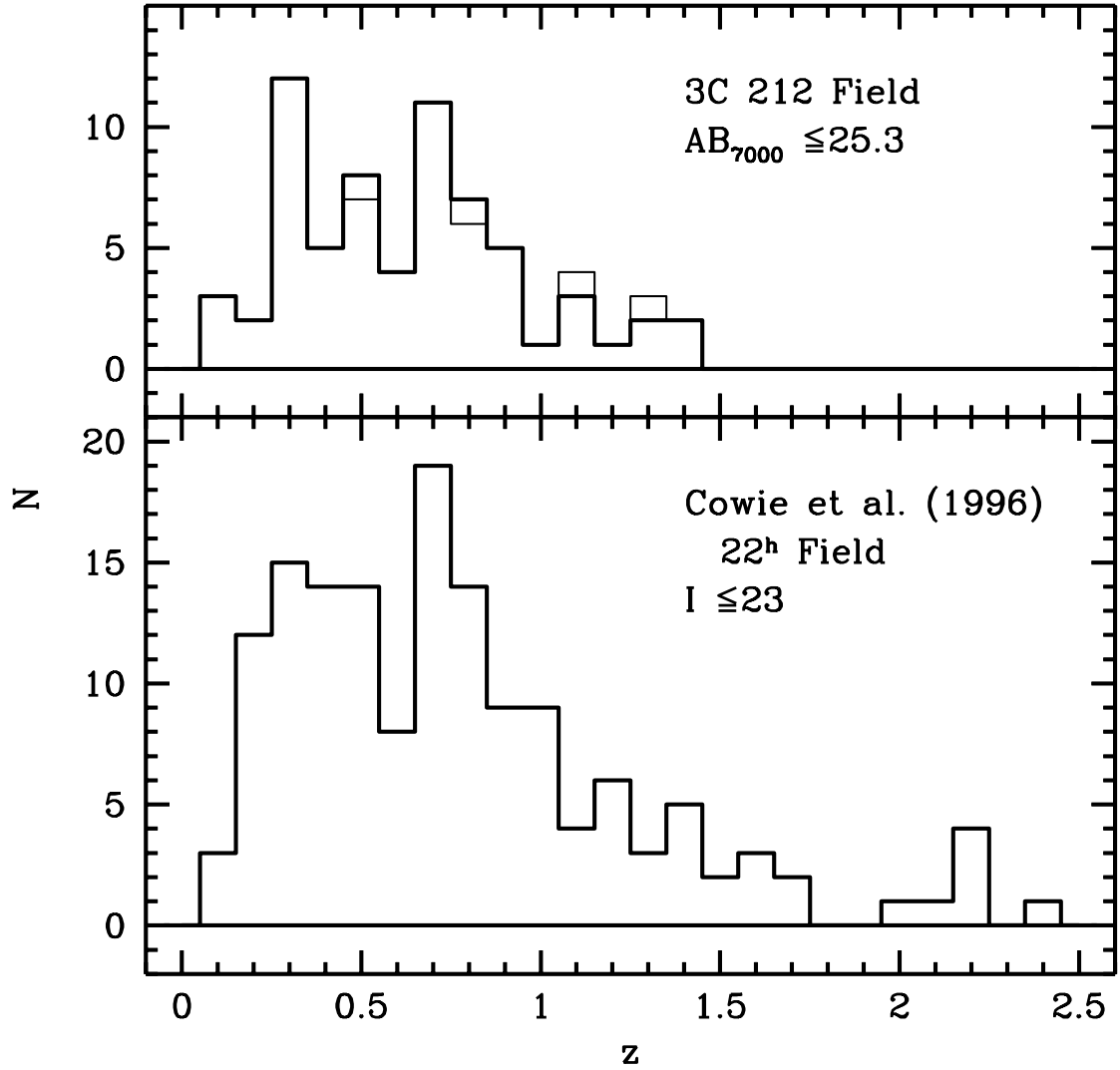}
\caption{Comparison of redshift distributions.  (a) {\it heavy line}---the
same redshift distribution of the 3C\,212 field as shown in Fig.\ 3, binned
to 0.1 intervals in $z$.  {\it lighter line}---redshift distribution after
a statistical correction for selection bias in the sample, as described in
the text.  (b)  Distribution of redshifts for the 22$^{\rm h}$ field of
Cowie \etal\ (1996).
}\label{zcomp}
\end{figure}

\end{document}